\pdfoutput=1

\documentclass[epj,nopacs]{svjour}
\usepackage{amssymb}
\usepackage{amsmath}
\usepackage{graphicx} 
\usepackage{color}
\usepackage{cite}
\usepackage{hyperref}

\begin{document}

\hugehead
\sloppy

\title {Multidimensional study of hadronization in nuclei}

\author{ 
The HERMES Collaboration \medskip \\
A.~Airapetian,$^{12,15}$
N.~Akopov,$^{26}$
Z.~Akopov,$^{5}$
E.C.~Aschenauer,$^{6,}$\footnote{Now at: Brookhaven National Laboratory, Upton, New York 11772-5000, USA}
W.~Augustyniak,$^{25}$
R.~Avakian,$^{26}$
A.~Avetissian,$^{26}$
E.~Avetisyan,$^{5}$
S.~Belostotski,$^{18}$
N.~Bianchi,$^{10}$
% ...
}

\institute{ 
$^1$Physics Division, Argonne National Laboratory, Argonne, Illinois 60439-4843, USA\\
$^2$Istituto Nazionale di Fisica Nucleare, Sezione di Bari, 70124 Bari, Italy\\
$^3$School of Physics, Peking University, Beijing 100871, China\\
$^4$Nuclear Physics Laboratory, University of Colorado, Boulder, Colorado 80309-0390, USA\\
$^5$DESY, 22603 Hamburg, Germany\\
$^6$DESY, 15738 Zeuthen, Germany\\
$^7$Joint Institute for Nuclear Research, 141980 Dubna, Russia\\
$^8$Physikalisches Institut, Universit\"at Erlangen-N\"urnberg, 91058 Erlangen, Germany\\
$^9$Istituto Nazionale di Fisica Nucleare, Sezione di Ferrara and Dipartimento di Fisica, Universit\`a di Ferrara, 44100 Ferrara, Italy\\
$^{10}$Istituto Nazionale di Fisica Nucleare, Laboratori Nazionali di Frascati, 00044 Frascati, Italy\\
$^{11}$Department of Subatomic and Radiation Physics, University of Gent, 9000 Gent, Belgium\\
$^{12}$Physikalisches Institut, Universit\"at Gie{\ss}en, 35392 Gie{\ss}en, Germany\\
$^{13}$SUPA, School of Physics and Astronomy, University of Glasgow, Glasgow G12 8QQ, United Kingdom\\
$^{14}$Department of Physics, University of Illinois, Urbana, Illinois 61801-3080, USA\\
$^{15}$Randall Laboratory of Physics, University of Michigan, Ann Arbor, Michigan 48109-1040, USA \\
$^{16}$Lebedev Physical Institute, 117924 Moscow, Russia\\
$^{17}$National Institute for Subatomic Physics (Nikhef), 1009 DB Amsterdam, The Netherlands\\
$^{18}$Petersburg Nuclear Physics Institute, Gatchina, 188300 Leningrad region, Russia\\
$^{19}$Institute for High Energy Physics, Protvino, 142281 Moscow region, Russia\\
$^{20}$Institut f\"ur Theoretische Physik, Universit\"at Regensburg, 93040 Regensburg, Germany\\
$^{21}$Istituto Nazionale di Fisica Nucleare, Sezione di Roma, Gruppo Collegato Sanit\`a and Istituto Superiore di Sanit\`a, 00161 Roma, Italy\\
$^{22}$TRIUMF, Vancouver, British Columbia V6T 2A3, Canada\\
$^{23}$Department of Physics, Tokyo Institute of Technology, Tokyo 152, Japan\\
$^{24}$Department of Physics and Astronomy, VU University, 1081 HV Amsterdam, The Netherlands\\
$^{25}$National Centre for Nuclear Research, 00-689 Warsaw, Poland\\
$^{26}$Yerevan Physics Institute, 375036 Yerevan, Armenia\\
} 

\date{Received: \today / Revised version:}

\authorrunning{The HERMES Collaboration}

\abstract{ Hadron multiplicities in semi-inclusive deep-inelastic
  scattering were measured on neon, krypton, and xenon targets relative
  to deuterium at an electron(positron)-beam energy of 27.6\,GeV at HERMES.
  These ratios were determined as a function of the virtual-photon
  energy $\nu$, its virtuality $Q^2$, the fractional hadron energy $z$
  and the transverse hadron momentum with respect to the
  virtual-photon direction $p_t$.  Dependences were analysed separately for
  positively and negatively charged pions and kaons as well as protons
  and antiprotons in a two-dimensional representation.  Compared to
  the one-dimensional dependences, some new features were observed.
  In particular, when $z>0.4$ positive kaons do not show the strong monotonic
  rise of the multiplicity ratio with $\nu$ as exhibited by pions and negative kaons.
  Protons were found to behave very differently from the other
  hadrons.  }

\maketitle

\section{Introduction}

The process of quark fragmentation and hadronization can be
investigated by measuring hadron production in semi-inclusive
deep-inelastic scattering of leptons from nuclei of various sizes.  As
typical hadronization lengths are of the order of the size of a
nucleus, the nuclei act as scale probes of the underlying
hadronization mechanism, {\it i.e.}, cross sections are expected to be
sensitive to whether---in a semi-classical picture---the 
hadronization occurs within or outside the
nucleus. Thus, the space (time) development of hadronization can be
investigated.  Such experiments were performed by the
SLAC~\cite{osborn}, EMC~\cite{ashm} and E665~\cite{adams}
collaborations. Recently, more precise data were collected and
analysed by the HERMES~\cite{herm1,herm2,herm3,herm4} and
CLAS~\cite{CLAS} collaborations.  The HERMES and 
the preliminary CLAS data are complementary in that the latter cover 
values of the virtual-photon energy $\nu$ below 4\,GeV.  Compared to 
hadronic probes, the use of leptonic probes has the advantage that
initial-state interactions can be neglected and that the energy and
momentum transferred to the struck parton are well determined by the
measured kinematic properties of the scattered lepton in the final
state.  The results of such studies of the hadronization in nuclei,
{\it i.e.}, cold nuclear matter, are expected to be useful for understanding
the fundamental aspects of hadronization in the framework of 
quantum chromodynamics, as well as an input
to the calculation of nuclear parton distribution functions (see, {\it e.g.}, 
ref.~\cite{nPDF}) and for the interpretation of jet-quenching and
parton energy-loss phenomena in ultra-relativistic heavy-ion
collisions (hot nuclear matter)~\cite{RHIC1,RHIC2,RHIC3,RHIC4,ATLAS,CMS,ALICE}.
 
The experimental data with leptonic probes are presented as the ratio
of the hadron multiplicities observed in the scattering on a nucleus
($A$) to those on the deuteron ($D$):

\begin{equation}
R_A^h(\nu,Q^2,z,p_t^2)=
\frac{\displaystyle \left(\frac{N^h(\nu,Q^2,z,p_t^2)}{N^e(\nu,Q^2)}\right)_{A}}
{\displaystyle \left(\frac{N^h(\nu,Q^2,z,p_t^2)}{N^e(\nu,Q^2)}\right)_{D}} \quad ,
\label{eq:test}
\end{equation}

\noindent where $N^h$ is the number of semi-inclusive hadrons in a
given ($\nu,Q^2,z,p_t^2$) bin and $N^e$ the number of inclusive
deep-inelastic scattering leptons in the same ($\nu,Q^2$) bin.  This
ratio depends on leptonic variables: the energy $\nu$ of the virtual photon
in the laboratory frame and its squared four-momentum $-Q^2$; and on
hadronic variables: the fraction $z$ of the virtual-photon energy
carried by the hadron and the square of the hadron momentum component
$p_{t}^2$ transverse to the virtual-photon direction.  In principle,
the ratio also depends on the azimuthal angle $\phi$ between the
lepton-scattering plane and the hadron-production plane.  
In the present measurement no dependence of $R_A^h$ on
$\phi$ was observed within the statistical accuracy. As a consequence
the integration over $\phi$ was performed.
 
In all previous publications results for $R_A^h$ were shown as a
function of one variable only (one-dimensional dependences) except
one case where a two-dimensional dependence was extracted for a
combined sample of charged pions~\cite{herm4}.  In the following, data
for $R_A^h$ for neon (Ne), krypton (Kr), and xenon (Xe)
are presented in a two-dimensional form for positively and negatively
charged pions and kaons, and for protons and antiprotons separately.
The two-dimensional representation consists in a fine binning in one
variable and a coarser binning in another variable.  The other
variables are integrated over within the acceptance of the experiment.
This allows the dependences to be studied in more detail, while
keeping the statistical uncertainties at moderate levels, at least for
pions, positive kaons and protons.  Some of the most prominent
features of the obtained results are presented and discussed. The full
set of results is available in a database~\cite{tdd}.

The wealth of theoretical model calculations and
studies~\cite{Accardi:2009qv, bialas1, bialas2, czyz, accardi1,
  accardi2, kop1, falter1, falter2, akopov1, akopov2, akopov3, wang1, wang2,
  wang3, arleo, gall, Sass} reflect the strong interest of the
community in hadron-multiplicities on nuclei, as they provide
information on the space (time) structure of the hadronization
process.  It is beyond the scope of this paper to compare the results
of the various models with the data. The one-dimensional results
published have already been instrumental in distinguishing between
some models~\cite{herm4}.  It is expected that the two-dimensional results
presented here will further help discriminating between
models.

\section{Experiment and data analysis}

The measurements were performed with the HERMES
spectrometer~\cite{herm0} using 27.6\,GeV positron and electron beams
stored in HERA at DESY.  Data were collected during 1999, 2000, 2004
and 2005 with gaseous targets of deuterium, neon, krypton and xenon.

The experimental set-up and data analysis are described in detail in
ref.~\cite{herm4}. Here, only a brief summary and update is given.
The identification of charged hadrons was accomplished using information
from the dual-radiator ring-imaging $\check{\mathrm{C}}$erenkov
detector (RICH)~\cite{herm5}, which provided separation of pions,
kaons and (anti)protons in the momentum range between 2 and 15\,GeV.
Compared to the analysis described in ref.~\cite{herm4}, an
improved hadron-identification algorithm was used, which is based on
a collective assignment of a set of identities to all particles
detected in the event, accounting for the correlations among their
probabilities~\cite{EVT,EVT2}. The differences compared to the results
obtained with the simpler approach neglecting such correlations were 
found to be minor and within the quoted
systematic uncertainties for all particles.

The scattered leptons were selected using the following kinematic
conditions: $\nu = 4.0 - 23.5$\,GeV (the upper bound corresponds to
$y=\nu/E < 0.85$), $Q^2 > 1$\,GeV$^2$, $W^2 > 4$\,GeV$^2$, where $E$
is the beam energy and $W$ is the invariant mass of the photon-nucleon
system. The constraints on $y$ and $W^2$ were applied in order to limit the
magnitude of radiative corrections and to suppress events originating 
from nucleon resonances, respectively.  The kinematic
constraints imposed on the selected hadrons were: $p_h = 2 - 15$\,GeV,
$z > 0.2$ and $x_F > 0$, where $p_h$ is the hadron momentum and the
Feynman variable $x_F$ is defined as the ratio of the longitudinal momentum
transferred to the hadron in the photon-nucleon centre-of-mass system 
to its maximum possible value.  
Together, the constraints on $z$ and $x_F$ reduce 
contributions from the target fragmentation region.

From the data, the hadron multiplicity ratios $R_A^{h}$ were
determined for each hadron type and target.  Radiative corrections
were applied following the scheme described in refs.~\cite{herm4, RC,
  RC1, RC2, RC3}, using average values of $\nu$ and $Q^2$ for each
kinematic bin in the analysis. The corrections remain below 7\% in all
bins.  Acceptance effects were studied in Monte-Carlo simulations
using an experimentally motivated parametrisation of $R_A^{h}$.  They
were found to be small compared to other uncertainties in all but the
lowest bin in $\nu$.  The differences between the parametrised and
reconstructed values were used to estimate the systematic uncertainty
due to the restricted acceptance for each hadron type.

Uncertainties in the knowledge of radiative processes (up to 2\%) and
half of the observed maximal differences between results for $R_A^{h}$
from different data-taking periods were taken together as overall
scale uncertainties.\footnote{In order to reduce effects from
  statistical fluctuations larger ranges of acceptance were integrated
  for these studies.  However, it was verified that those effects were
  not generated in certain kinematic ranges only.}  The total scale
uncertainties are 3\%, 5\%, 4\%, and 10\% for pions, kaons, protons
and antiprotons, respectively.

The uncertainties due to the hadron identification were estimated to
be up to 0.5\% for charged pions, up to 1.5\% for kaons and protons,
and up to 4\% for antiprotons. Those due to acceptance effects were
6\% for pions, 3\% for kaons, and 7\% for protons and antiprotons in
the first $\nu$ bin, and less than 2\% for any hadron in any other
bin. Effects due to the contamination from diffractive $\rho^0$
meson production were estimated to be at most 4 and 7\% for positive
and negative pions, respectively.  (For details see
ref.~\cite{herm4}.)
These uncertainties were added in quadrature separately for each data
point to yield systematic bin-to-bin uncertainties.
Those were subsequently added in quadrature to the statistical
uncertainties and plotted as total uncertainties.

\section{Results and discussion}

\begin{figure*}[th]
\center{
\includegraphics[width=0.485\textwidth]{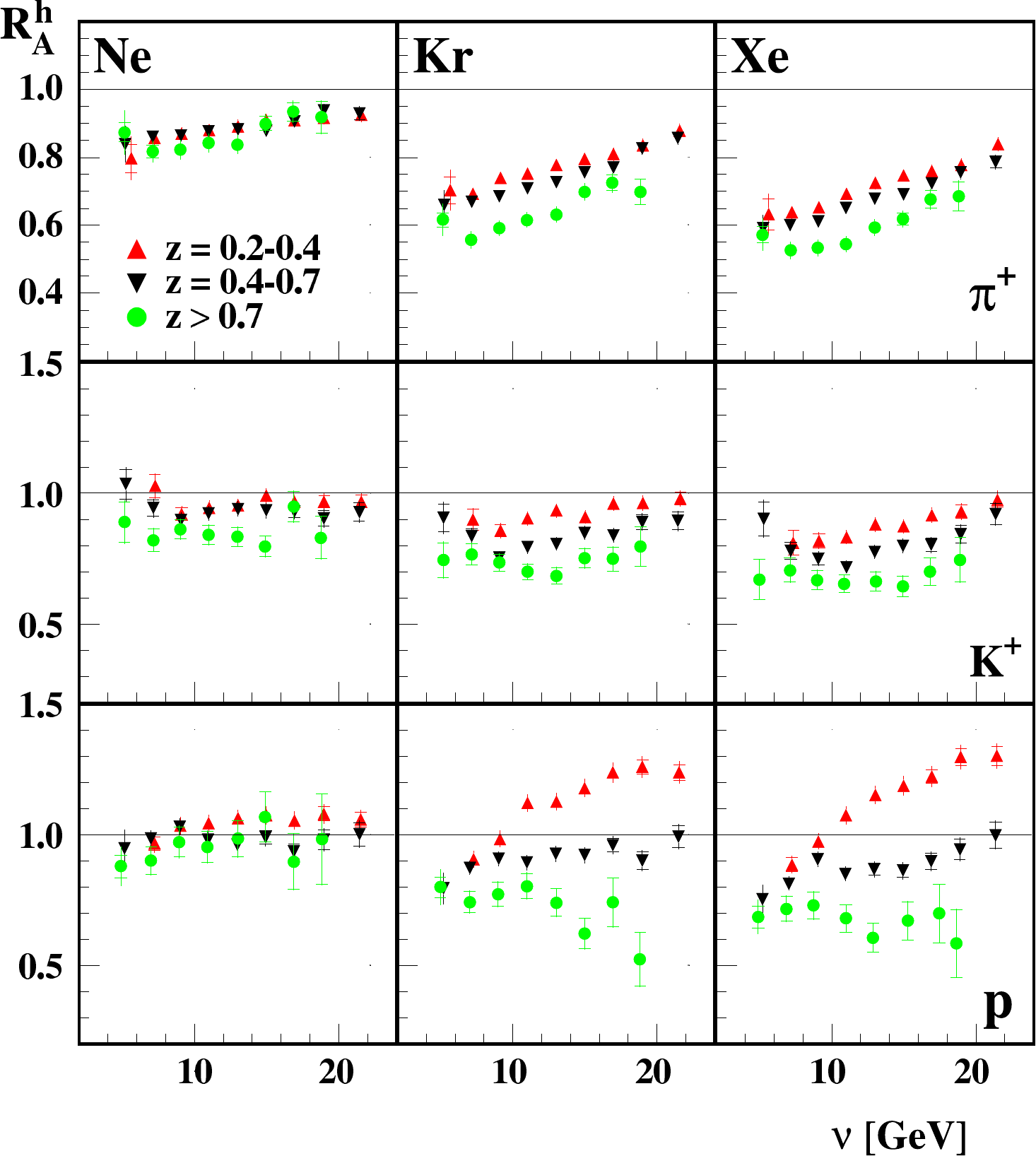}
\hspace*{2mm}
\includegraphics[width=0.485\textwidth]{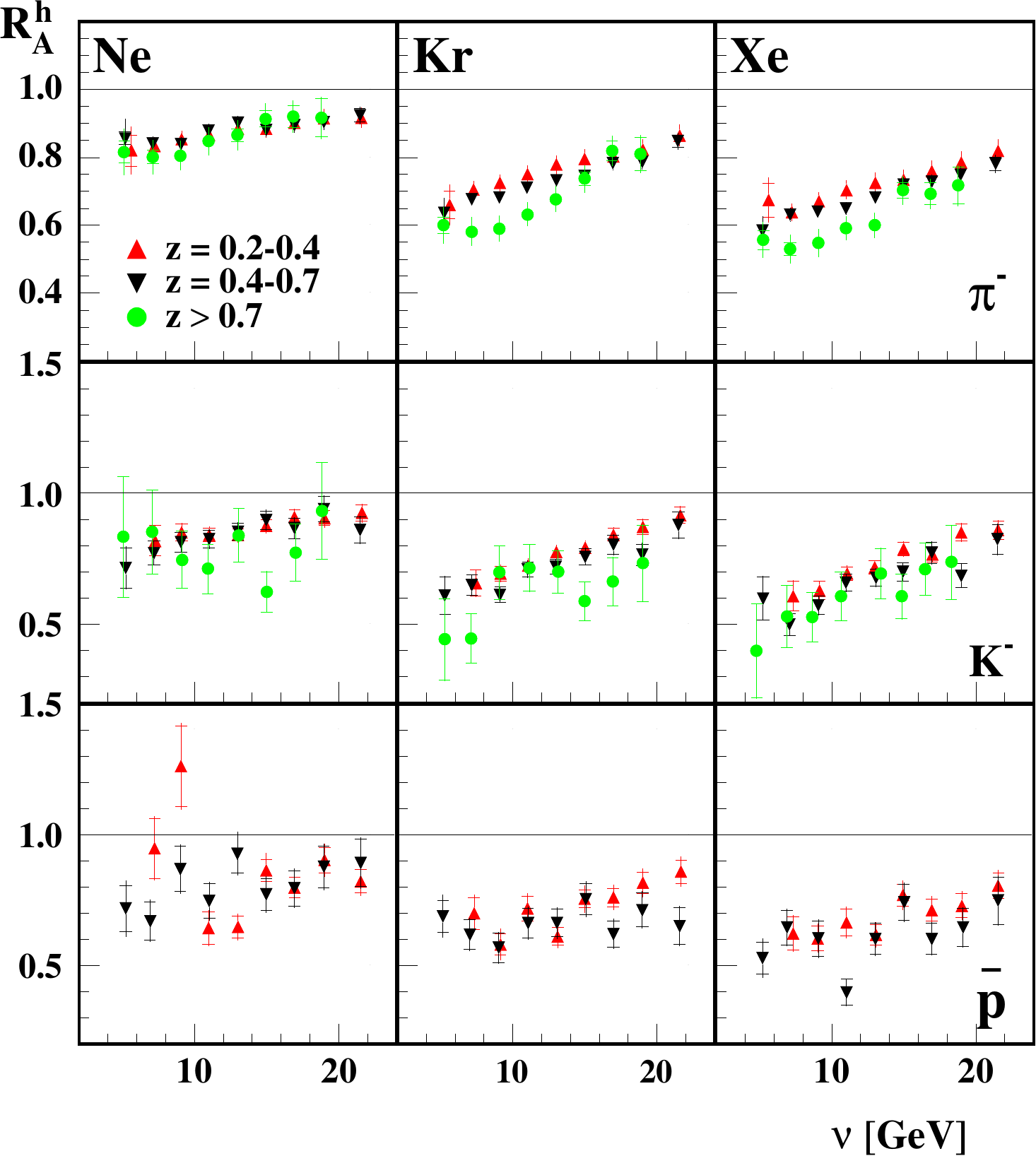}
\caption{Dependence of $R_{A}^h$ on $\nu$ for positively and
  negatively charged hadrons for three slices in $z$ as indicated in
  the legend.  The inner and outer error bars indicate the statistical
  and total uncertainties, respectively.  For the latter the
  statistical and systematic bin-to-bin uncertainties were added in
  quadrature.  In addition, scale uncertainties of 3\%, 5\%, 4\%, and
  10\% are to be considered for pions, kaons, protons and
  antiprotons, respectively.}
\label{fig:nu_z}
}
\end{figure*}

The results for the multiplicity ratio $R_A^{h}$ are presented using a
fine binning in one of the variables, a coarser binning (called slice)
in a second variable, and integrating over the remaining variables
within the acceptance of the experiment.  The following slices were
used: $4-12$, $12-17$, and $17-23.5$\,GeV for $\nu$; $0.2-0.4$,
$0.4-0.7$, and $> 0.7$ for $z$; and $\le 0.4$, $0.4-0.7$, and $>
0.7$\,GeV$^2$ in the case of $p_t^2$.  The dependence on $Q^2$ was
investigated, but as it turned out to be weak, no dependences with
slices in $Q^2$ were produced.  In the following, dependences that
show salient features are discussed.  In the presentation of the data,
bins based on fewer than 10 events were omitted because the large
statistical uncertainty would preclude useful conclusions.

The dependence of $R_{A}^h$ on $\nu$ for three slices in $z$ is shown
in fig.~\ref{fig:nu_z}.  For pions and K$^-$, a global trend of steady
increase of $R_A^h$ with increasing values of $\nu$ was observed.
Such a behaviour is explained in fragmentation models as resulting
from Lorentz dilation and/or a shift in the argument $z$ of the
relevant fragmentation function~\cite{Accardi:2009qv}.  However, 
at the highest $z$ range there is an indication for
a flattening out (and possibly a reversal of
this trend) at low $\nu$  for $\pi^+$ and $\pi^-$
independently, which is not explained by these mechanisms.

The behaviour of $R_{A}^h$ for $\mathrm{K}^+$ was found to be more
complicated.  For krypton and xenon there is a clear increase of
$R_{A}^{\mathrm{K}^+}$ with $\nu$ for the lowest $z$-slice, but at
larger values of $z$ the behaviour is flatter. In contrast, the
results for $R_{A}^h$ for $\mathrm{K}^-$ resemble those for pions.
For antiprotons, the $\nu$-dependence was found to be weak with a
slightly positive slope, but the statistical accuracy of the results
is too limited to draw definite conclusions. The neon data show
similar but less pronounced trends, which was a common observation in
all distributions under study. This is not unexpected due to the
smaller size of the nucleus of neon compared to krypton and xenon.

The results for protons differ significantly from those for the other
hadrons.  For the heavy nuclei, $R_A^{\mathrm{p}}$ behaves very
differently for the three $z$-slices, considerably exceeding unity at
higher $\nu$ for the lowest $z$-slice.  Part of the explanation may be
the following.  Unlike the other hadrons, protons are present already
in the target nucleus.  Therefore, apart from hadronization, residual
protons can also result from reactions in the final state (final-state
interactions), whereby a proton is knocked out of the nucleus. Those
protons will preferably be emitted with low energy.  This could lead
to an energy dependence which, in conjunction with other kinematic
factors, leads to the observed non-trivial behaviour.

\begin{figure*}[bth]
\center{
\includegraphics[width=0.485\textwidth]{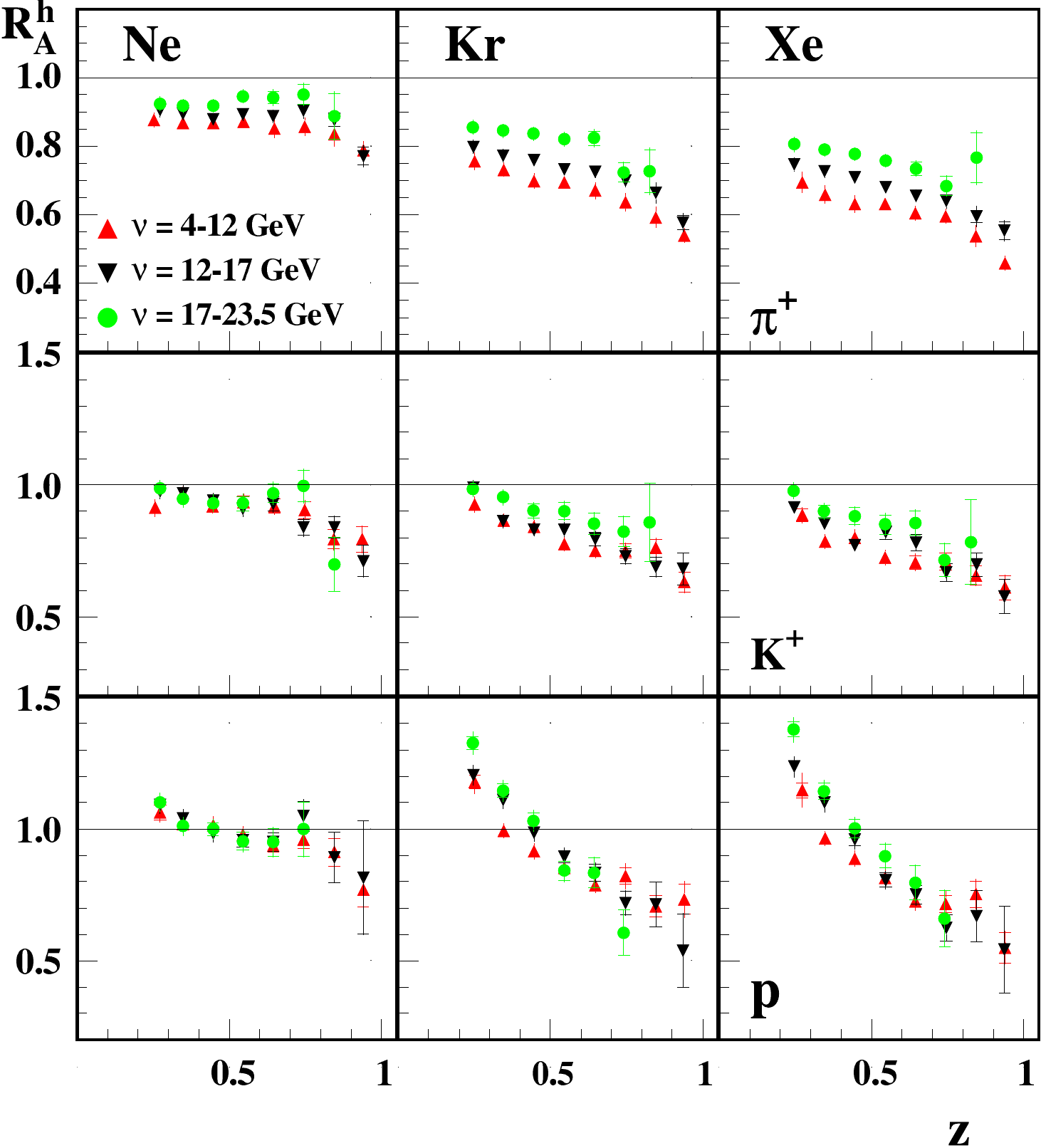}
\hspace*{2mm}
\includegraphics[width=0.485\textwidth]{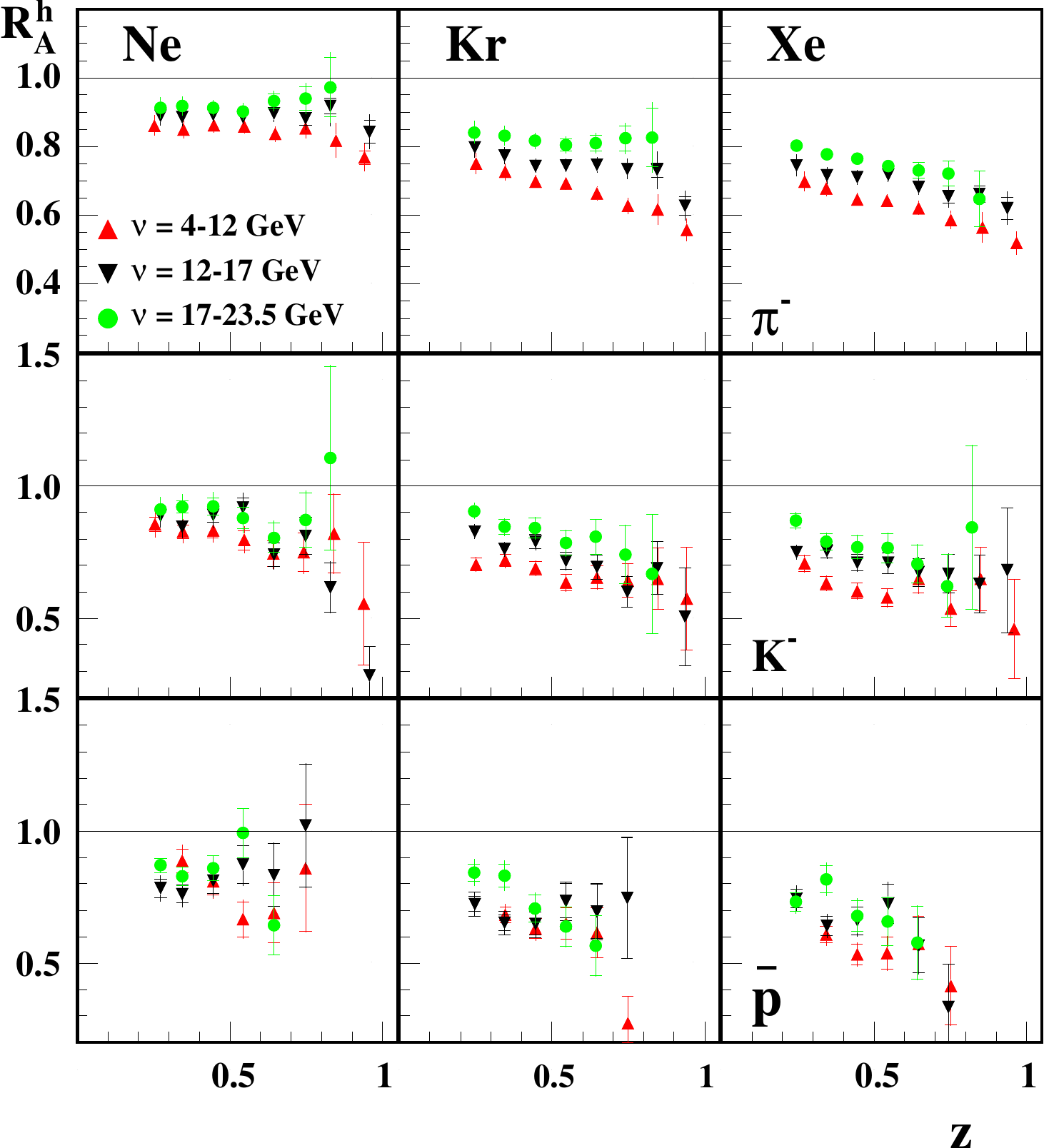}
\caption{Dependence of $R_{A}^h$ on $z$ for positively and
  negatively charged hadrons for three slices in $\nu$ as indicated
  in the legend.  Uncertainties are
  shown as in fig.~\ref{fig:nu_z}. }
\label{fig:z_nu}
}
\end{figure*}

The dependence of $R_{A}^h$ on $z$ for three\footnote{As the combined
  dependence on $\nu$ and $z$ is crucial for various model
  calculations, the results as a function of $z$ are also given for
  five slices in $\nu$ in ref.~\cite{tdd}.}  slices in $\nu$ is shown
in fig.~\ref{fig:z_nu}.  A slight change of the $z$ dependence when
varying the $\nu$ range was observed for the $\pi^+$ and $\pi^-$
distributions.  This has been observed already in ref.~\cite{herm4}
for the combined pion sample and we refer to that paper for the
discussion.  The results on krypton and xenon for protons show a very
strong dependence on $z$, the value of $R_A^{\mathrm{p}}$ exceeding
unity in all $\nu$ ranges at low $z$.  This supports the assumption
that at low values of $z$ there is a sizable contribution of
final-state interactions.  A similar, but smaller effect was seen for
K$^+$, as $R_A^{\mathrm{K}^+}$ increases to almost unity, while
$R_A^{\mathrm{K}^-}$ remains well below unity. This suggests that
interactions play a role for K$^+$ production in which a proton in the
target nucleus is transformed into a K$^+$ $\mathrm{\Lambda}$ pair
while the analogous process for K$^-$ production is suppressed due to
the quark content of the K$^-$~\cite{kop1}.

\begin{figure}[!ht]
\center{
\includegraphics[width=\columnwidth]{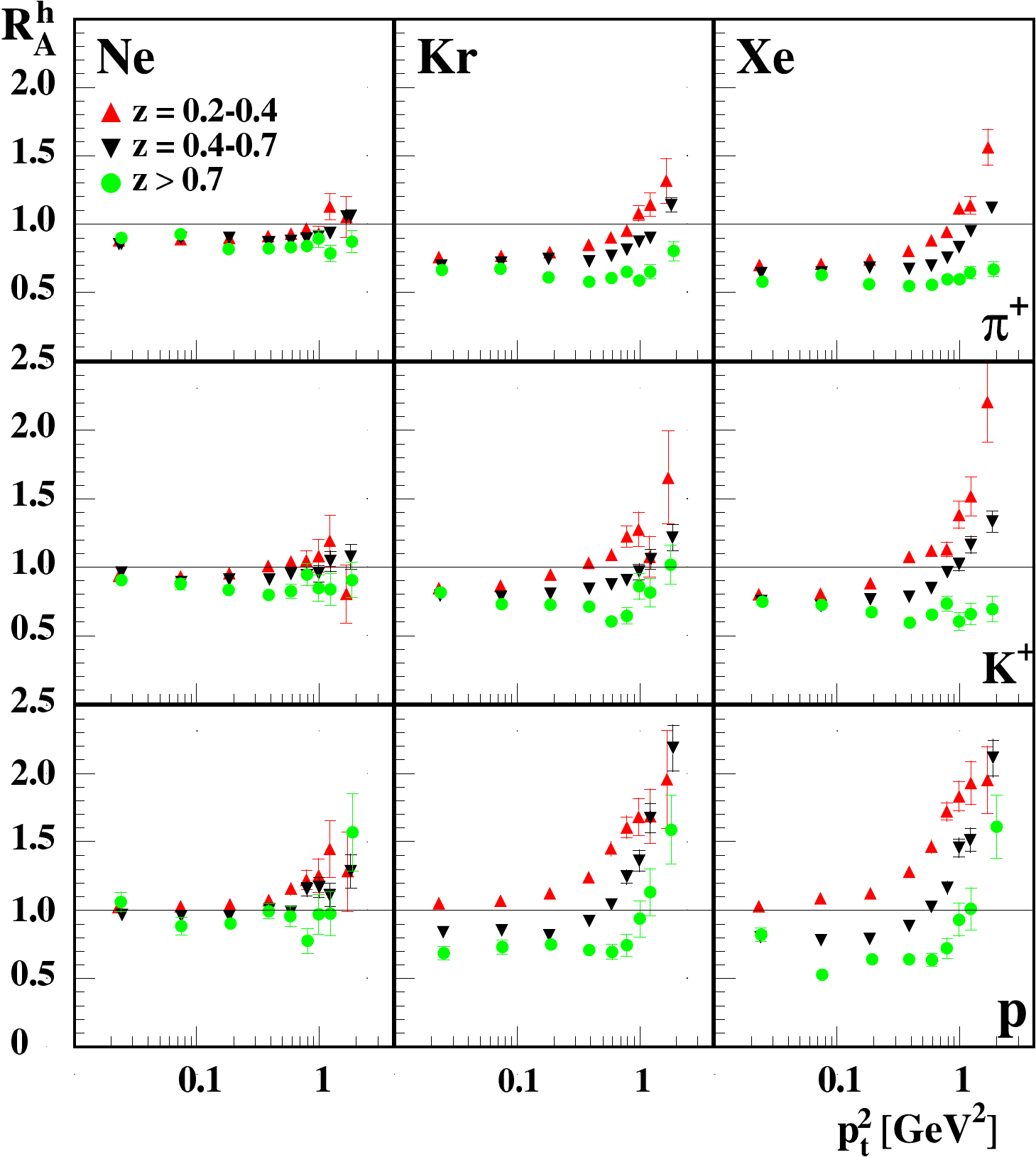}
\caption{Dependence of $R_{A}^h$ on $p_{t}^{2}$ for positively charged
  hadrons for three slices in $z$ as indicated in the legend.
  Uncertainties are shown as in fig.~\ref{fig:nu_z}.}
\label{fig:pt_z}
}
\end{figure}

Figure~\ref{fig:pt_z} shows the dependence of $R_{A}^h$ on $p_t^2$ for
three slices in $z$ for positively charged hadrons.  The behaviour of
$R_{A}^h$ for $\pi^-$ (not shown) was found to be the same as that
for $\pi^+$ within statistical uncertainties.  The rise at high
$p_t^2$ suggests a broadening of the $p_t$ distribution~\cite{kop1}.
Such a broadening could result from an interaction of the struck quark
with the nuclear environment before the final hadron is produced
and/or from interactions of the produced hadron within the nucleus.  A
detailed analysis and discussion of the HERMES data for pions and
K$^+$ particles in terms of $p_t$-broadening has been presented in
ref.~\cite{ptbroad}.  Interesting to note is that in the highest
$z$-slice $R_{A}^h$ for pions and $\mathrm{K}^+$ becomes independent
of $p_t^2$ within statistical uncertainties, while for protons a
significant rise is observed at high $p_t^2$.  For $\mathrm{K}^-$ and
antiprotons (neither are shown) limited statistics preclude any
definite conclusion.  In the intermediate $z$-range protons also show
a much stronger rise with $p_t^2$ compared to pions and kaons in the
respective ranges.  This is consistent with a large contribution of
final-state interactions in the case of protons.

\begin{figure*}[th]
\center{
\includegraphics[width=0.48\textwidth]{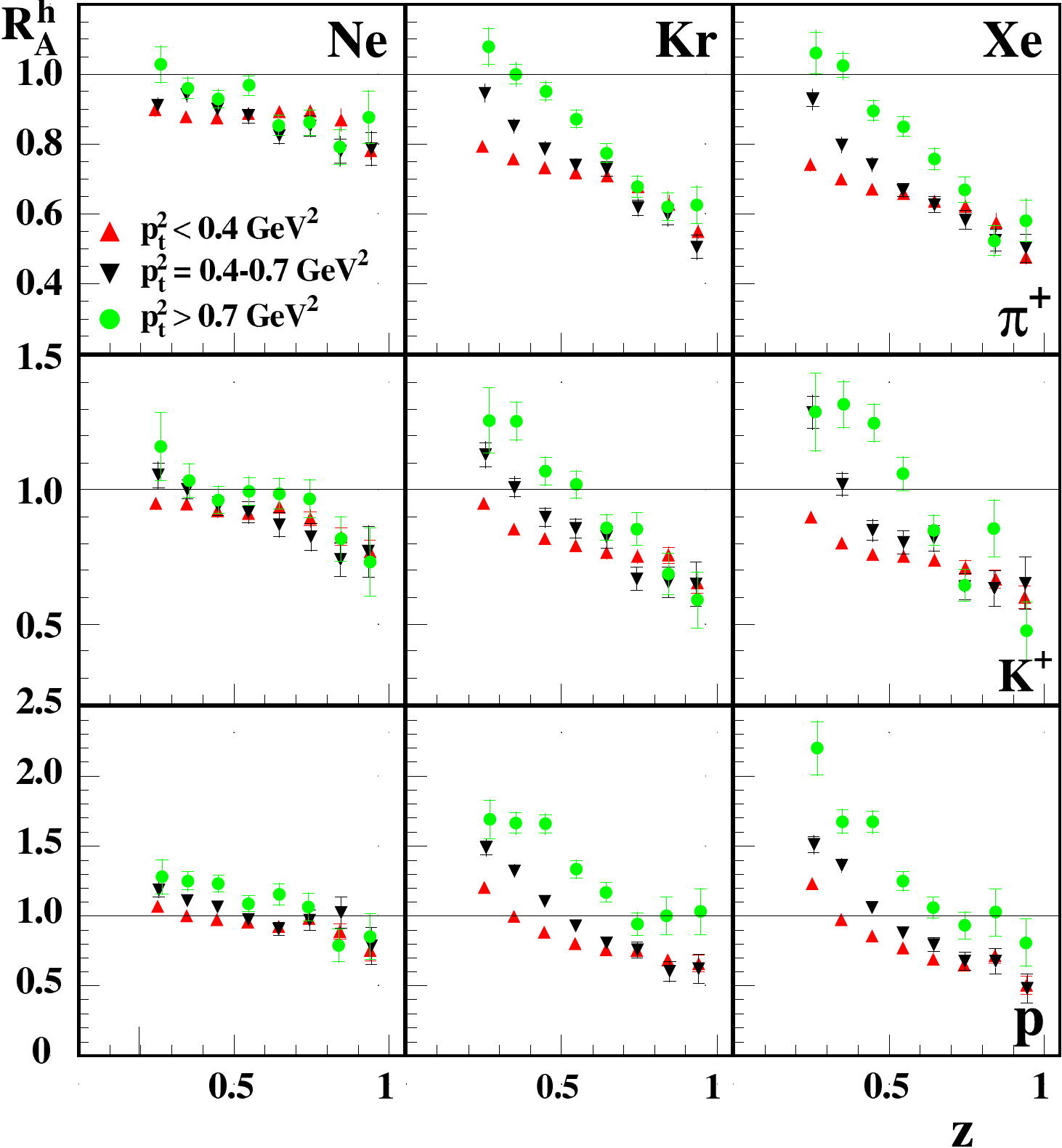}
\hspace*{2mm}
\includegraphics[width=0.48\textwidth]{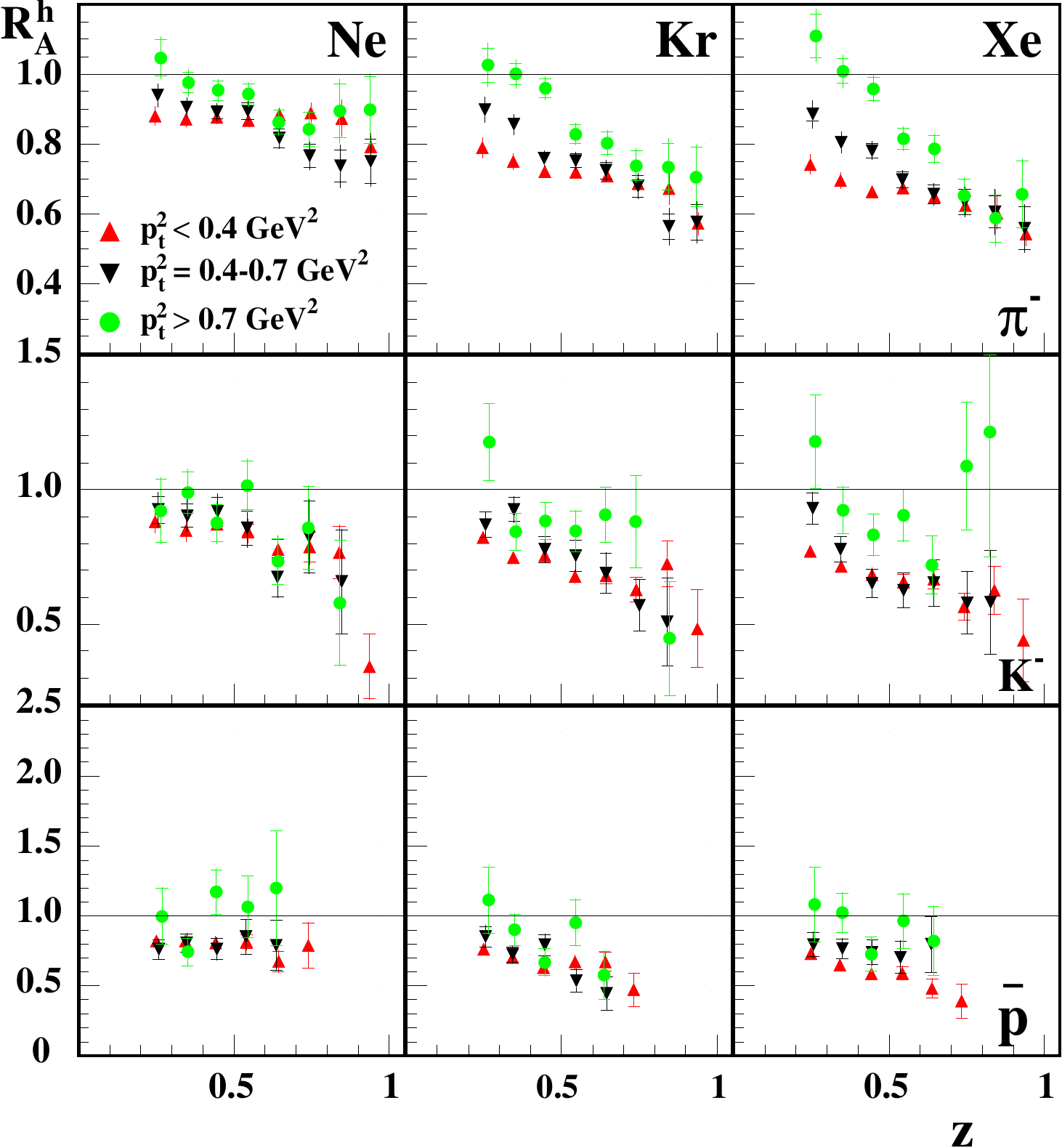}
\caption{Dependence of $R_{A}^h$ on $z$ for positively and negatively
  charged hadrons for three slices in $p_{t}^2$ as indicated in the
  legend.  Uncertainties are shown as in
  fig.~\ref{fig:nu_z}.}
\label{fig:z_pt}
}
\end{figure*}

In fig.~\ref{fig:z_pt}, the variation of the $p_t^2$-dependence with
$z$ is presented in a different way by showing the dependence of
$R_{A}^h$ on $z$ for three slices in $p_t^2$.  The global decrease of
$R_{A}^h$ with $z$ was already observed in fig.~\ref{fig:z_nu}.  This
dependence of $R_{A}^h$ on $z$ turns out to be stronger at higher
values of $p_t^2$, an effect that is emphasised at larger target
mass. At high $z$, the dependence on $p_t^2$ disappears for $\pi^+$,
$\pi^-$, and $\mathrm{K}^+$, as has already been seen in
fig.~\ref{fig:pt_z}.  This lack of nuclear broadening of the $p_t$
distribution in the limit of instantaneous hadronization, {\it i.e.}, 
before the struck parton has lost any energy, has been interpreted
in terms of broadening arising from partonic processes~\cite{kop1}.
For protons, a similar, but much stronger dependence of the slope on
$p_t^2$ was observed, with $R_A^{\mathrm{p}}$ increasing far above
unity at low $z$. This has been discussed in relation to
fig.~\ref{fig:z_nu} as being an indication of final-state
interactions.  The fact that the values of $R_{A}^h$ for K$^+$ are in
between those for pions and protons suggests that, in addition to
fragmentation, again final-state interactions play a role here.  The
large uncertainties of $R_{A}^h$ for $\mathrm{K}^-$ and antiprotons
preclude any particular conclusion in those cases.

\section{Conclusions}

Two-dimensional kinematic dependences have been presented for the
multiplicity ratio $R_A^{h}$ for identified $\pi^+$, $\pi^-$,
$\mathrm{K}^+$, $\mathrm{K}^-$, protons and antiprotons, measured in
semi-inclusive deep-inelastic scattering of electrons and positrons from
either neon, krypton, or xenon relative to deuterium.
These two-dimensional distributions provide detailed information,
which in some cases is not accessible in the one-dimensional
distributions (in which all kinematic variables except one are
integrated over, as has been traditionally done).

The behaviour of $R_A^{h}$ for $\pi^+$ and $\pi^-$ was found to be
the same within the experimental uncertainties and is globally
characterised by an increase of $R_A^{h}$ with the total energy
transfer $\nu$ and a decrease with the fractional energy $z$ of the
produced hadron.  Negatively charged kaons behave similarly to pions,
while the dependence of $R_A^{\mathrm{K}^+}$ for positively charged kaons on $\nu$
changes depending on the value of $z$, possibly due to final-state interactions. 
Protons behave very differently from the other hadrons, especially in the
$\nu$-distribution for different values of $z$. This may be explained
by a sizable contribution of final-state interactions, such as
knock-out processes, in addition to the fragmentation process.  These
new detailed data are expected to be an essential ingredient for
constraining models of hadronization and, hence, improving our
understanding of hadron formation.

\begin{acknowledgement}
We gratefully acknowledge the DESY management for its support and the staff
at DESY and the collaborating institutions for their significant effort.
This work was supported by 
the Ministry of Economy and the Ministry of Education and Science of Armenia;
the FWO-Flanders and IWT, Belgium;
the Natural Sciences and Engineering Research Council of Canada;
the National Natural Science Foundation of China;
the Alexander von Humboldt Stiftung;
the German Bundesministerium f\"ur Bildung und Forschung (BMBF);
the Deutsche Forschungsgemeinschaft (DFG);
the Italian Istituto Nazionale di Fisica Nucleare (INFN);
the MEXT, JSPS, and G-COE of Japan;
the Dutch Foundation for Fundamenteel Onderzoek der Materie (FOM);
the Russian Academy of Science and the Russian Federal Agency for 
Science and Innovations;
the U.K.~Engineering and Physical Sciences Research Council, 
the Science and Technology Facilities Council,
and the Scottish Universities Physics Alliance;
the Basque Foundation for Science (IKERBASQUE);
the U.S.~Department of Energy (DOE) and the National Science Foundation (NSF);
and the European Community Research Infrastructure Integrating Activity
under the FP7 ``Study of strongly interacting matter'' (HadronPhysics2, Grant
Agreement number 227431).
\end{acknowledgement}

\end{document}